\documentclass[showpacs,preprintnumbers,amsmath,amssymb,aps]{revtex4}

\begin{document}

\title{Inflaton in anisotropic higher derivative gravity}
\author{
W. F. Kao\thanks{%
gore@mail.nctu.edu.tw} \\
Institute of Physics, Chiao Tung University, Hsinchu, Taiwan}

\begin{abstract}
Existence and stability analysis of the Kantowski-Sachs type
inflationary universe in a higher derivative scalar-tensor gravity
theory is studied in details. Isotropic de Sitter background
solution is shown to be stable against any anisotropic
perturbation during the inflationary era. Stability of the de
Sitter space in the post inflationary era can also be realized
with proper choice of coupling constants.
\end{abstract}

\maketitle

\section{Introduction}

Our universe is known to be homogeneous and isotropic to a very
high degree of precision \cite{data,cobe}. Such an universe can be
described by the well known Friedmann-Robertson-Walker (FRW)
metric \cite{book}. There had been, however, some cosmological
problems associated with the standard big bang model responsible
for the evolution of our present universe. Inflationary models
resolves many problems including the flatness, monopole, and
horizon problem\cite{inflation}.

Moreover, gravitational physics is expected to be different from
the Einstein-Hilbert models near the Planck scale
\cite{string,scale}. For example, quantum gravity or string
corrections indicate that higher derivative terms could have some
interesting cosmological applicastions\cite{string}. In addition,
higher derivative terms also arise as the quantum corrections of
the matter fields \cite{green}. Therefore, the possibility of
deriving inflation from higher derivative corrections have been an
focus of research interest \cite{jb1,jb2,Kanti99,dm95,green}.

For example, a general analysis on the stability conditions of
gravity theories has been shown to be useful in the search of
physical models compatible with our physical universe. In
particular, the stability condition for a variety of pure gravity
theories as a potential candidate of inflationary universe in the
flat Friedmann-Robertson-Walker (FRW) space is derived in Ref.
\cite{dm95,kp91,kpz99}.

In addition, the highly isotropic universe should be stable
against anisotropic perturbation so that the isotropic FRW space
can remain a stable final state. The initial state of our universe
could also be anisotropic before it becomes isotropic at later
stage of the evolution. Nonetheless, it is interesting itself to
study the stability analysis of the anisotropic space during the
post-inflationary epoch. For example, it is shown that anisotropic
inflationary solution does exist for an NS-NS model with a metric,
a dilaton, and an axion field \cite{CHM01}. This inflationary
solution is also shown to be stable against small perturbations
\cite{ck01}. Similar stability analysis has also been studied for
a various of interesting models\cite{abel}.

Recently, there are also growing interests in the study of
Kantowski-Sachs (KS) type anisotropic spaces\cite{BPN,LH,NO}. We
will hence try to study the problem of existence and stability
associated with the inflationary solution with an de Sitter final
state. In particular, we will study the effects of the higher
derivative terms in Kantowski-Sachs spaces. In fact, a large class
of pure gravity models and induced gravity models with
inflationary KS$/$FRW solutions was presented in Ref.
\cite{kao06}. Any KS type solution leading itself to an asymptotic
FRW final state will be referred to as the KS$/$FRW solution in
this paper for convenience.

It is shown that the stability of the de Sitter background space
is closely related to the choice of the coupling
constants\cite{kao06}. Indeed, the pure gravity model given below:
\begin{equation}
{\cal L}_g = -R - \alpha R^2 - \beta R^\mu_\nu R^\nu_\mu + \gamma
R^{\mu \nu}_{\;\;\; \beta \gamma} \, R^{\beta \gamma}_{\;\;\;
\sigma \rho} \, R^{\sigma \rho}_{\;\;\; \mu \nu}
\end{equation}
admits an inflationary solution with a constant Hubble parameter
given by $H_0^4 = 1/4 \gamma$. This will requires that $\gamma
>0$. Here $\alpha$, $\beta$, and $\gamma$ are coupling constants.
This shows that (a) the $\gamma$ factor determines the scale of
the inflation characterized by the Hubble parameter $H_0$ and (b)
$\alpha$ and $\beta$ factors are irrelevant to the scale $H_0$ in
the de Sitter phase. The quadratic terms are, however, important
to the stability of the de Sitter phase.

Indeed, perturbing the KS type metric with $H_i \to H_0 +\delta
H_i$, we can show that
\begin{equation}
\delta H_i = c_i \exp [{-3H_0t \over 2} (1+ \delta_1) ] + d_i \exp
[{-3H_0t \over 2} (1- \delta_1) ]
\end{equation}
for
\begin{equation}
\delta_1 = \sqrt{1+ 8/[27-9 (6 \alpha+2 \beta)H_0^2 ] }
\end{equation}
and some arbitrary constants $c_i, d_i$ to be determined by the
initial perturbations. Here $H_i\equiv \dot{a}_i/a_i$ with
$a_i(t)$ the scale factor in $i$-direction. We will describe the
notation shortly in section II. It is easy to see that any small
perturbation $\delta H_i$ will be stable against the de Sitter
background if both modes characterized by the exponents
\begin{equation}
\Delta_{\pm} \equiv - [3H_0t / 2] [1 \pm \delta_1]
\end{equation}
are all negative. This will happen if $\delta_1 < 1$. In such
case, the inflationary de Sitter space will remain a stable
background as the universe evolves.

It can be shown that the stability equation for the anisotropic KS
space and the stability equation for the isotropic FRW space in
the presence of the same inflationary de Sitter background turns
out to be identical \cite{dm95,kp91, kpz99}. Therefore, the
stability of isotropic perturbations also ensures the stability of
the anisotropic perturbations. The stability of the isotropic
perturbations for the FRW space is important for any physical
models. Unfortunately, inflationary models that are stable against
any isotropic perturbations will have problem with the graceful
exit process. Therefore, the pure gravity model may have troubles
dealing with the stability and exit mechanism all together.

Instead of the pure gravity theory, a slow rollover scalar field
may help resolving this problem. An inflationary de Sitter
solution in a scalar-tensor model is expected to have one stable
mode (against the perturbation in $\delta H_i$ direction) and one
unstable mode (against the perturbation in $\delta \phi$
direction). As a result, the inflationary era will come to an end
once the unstable mode takes over after a brief period of
inflationary expansion. Therefore, we propose to study the effect
of such theory.

In particular, we will show in this paper that the roles played by
the higher derivative terms are dramatically different in the
inflationary phase of our physical universe in both pure gravity
theory and scalar-tensor theory. First of all, third order term
will be shown to determine the expansion rate $H_0$ for the
inflationary de Sitter space. The quadratic terms will be shown to
have nothing to do with the expansion rate of the background de
Sitter space. They will however affect the stability condition of
the de Sitter phase. Their roles played in the existence and
stability condition of the evolution of the de Sitter space are
dramatically different.

\section{Non-redundant field equation and Bianchi identity in KS space}

Given the metric of the following form:
\begin{equation}
ds^2=- dt^2+c^2(t)dr^2 + a^2(t) ( d^2 \theta +f^2(\theta) d
\varphi^2)
\end{equation}
with $f(\theta)= (\theta, \sinh \theta, \sin \theta)$ denoting the
flat, open and close anisotropic space known as Kantowski-Sachs
type anisotropic spaces. More specifically, Bianchi I (BI),
III(BIII), and Kantowski-Sachs (KS) space corresponds to the flat,
open and closed model respectively. This metric can be rewritten
as
\begin{equation} \label{metric}
ds^2=- dt^2+a^2(t)({dr^2\over {1-kr^2}}+r^2d\theta ^2) + a_z^2(t)
dz^2
\end{equation}
with $r$, $\theta$, and $z$ read as the polar coordinates and $z$
coordinate for convenience and for easier comparison with the FRW
metric. Note that $k=0,1,-1$ stands for the flat, open and closed
universes similar to the FRW space.

Writing $H_{\mu \nu} \equiv G_{\mu \nu} -T_{\mu \nu}$, Einstein
equation can be written as $D_\mu H^{\mu \nu}=0$ incorporating the
Bianchi identity $D_\mu G^{\mu \nu}=0$ and the energy momentum
conservation $D_\mu T^{\mu \nu}=0$. Here $G^{\mu \nu}$ and $T^{\mu
\nu}$ represent the Einstein tensor and the energy momentum tensor
coupled to the system respectively. With the metric
(\ref{metric}), it can be shown that the $r$ component of the
equation $D_\mu H^{\mu \nu}=0$ implies that
\begin{equation}
H^r_{\; r}=H^\theta_{\;\theta}.
\end{equation}
This result also says that any matter coupled to the system has
the symmetric property $T^r_{\;r}=T^\theta_{\;\theta}$. In
addition, the equations $D_\mu H^{\mu \theta}=0$ and $D_\mu H^{\mu
z}=0$ both vanish identically for all kinds of energy momentum
tensors. More interesting information comes from the $t$ component
of this equation. It says:
\begin{equation}
(\partial_t + 3 H) H^t_{\; t} = 2 H_1 H^r_{\; r} +H_zH^z_{\; z}.
\end{equation}
This equation implies that (i) $H^t_{\; t}=0$ implies that
$H^r_{\; r}=H^z_{\; z}=0$ and (ii) $H^r_{\; r}=H^z_{\; z}=0$ only
implies $(\partial_t + 3 H) H^t_{\; t} =0$ instead of $H^t_t=0$.
Case (ii) can be solved to give $ H^t_{\; t} =$ constant $\times
\exp[-a^2a_z]$ which approaches zero when $a^2a_z \to \infty$. For
the anisotropic KS spaces, the metric contains two independent
variables $a$ and $a_z$. The Einstein field equations have,
however, three non-vanishing components: $H^t_{\; t} =0$, $H^r_{\;
r}=H^\theta_{\;\theta} =0$ and $H^z_{\; z} =0$. The Bianchi
identity implies that the $tt$ component is not redundant and will
hence be retained for complete analysis. Ignoring either one of
the $rr$ or $zz$ components will not affect the final result of
the system. In short, the $H^t_t=0$ equation, known as the
generalized Friedman equation, is a non-redundant field equation
as compared to the $H^r_r=0$ and $H^z_z=0$ equations.

In addition, restoring the $g_{tt}$ component $b^2(t)=1/B_1$ will
be helpful in deriving the non-redundant field equation associated
with $G_{tt}$ that will be shown shortly. More specifically, the
generalized KS metric will be written as:
\begin{equation}  \label{metricb}
ds^2=-b^2(t) dt^2+a^2(t)({dr^2\over {1-kr^2}}+r^2d\theta ^2) +
a_z^2(t) dz^2.
\end{equation}
In principle, the Lagrangian of the system can be reduced from a
functional of the metric $g_{\mu \nu}$, ${\cal L}(g_{\mu \nu})$,
to a simpler function of $a(t)$ and $a_z(t)$, namely $L(t) \equiv
a^2 a_z {\cal L}(g_{\mu \nu}(a(t), a_z(t)))$. The equation of
motion should be reconstructed from the variation of the reduced
Lagrangian $L(t)$ with respect to the variable $a$ and $a_z$. The
result is, however, incomplete because, the variation of $a$ and
$a_z$ are related to the variation of $g_{rr}$ and $g_{zz}$
respectively. The field equation from varying $g_{tt}$ can not be
derived without restoring the variable $b(t)$ in advance. This is
the motivation to introduce the metric (\ref{metricb}) such that
the reduced Lagrangian $L(t) \equiv ba^2 a_z {\cal L}(g_{\mu
\nu}(b(t),a(t), a_z(t)))$ retains the non-redundant information of
the $H^t_t=0$ equation. Non-redundant Friedmann equation can be
reproduced resetting $b=1$ after the variation of $b(t)$ has been
done.

After some algebra, all non-vanishing components of the curvature
tensor can be computed: \cite{kao06}
\begin{eqnarray}
R^{ti}_{\;\;\;tj} &=& [{1\over 2}\dot{B}_1H_i+B_1 ( \dot{H}_i+H^2_i) ] \delta^i_j ,\\
R^{ij}_{\;\;\;kl} &=& B_1H_iH_j \; \epsilon^{ijm}\epsilon_{klm}+{k
\over a^2} \epsilon^{ijz}\epsilon_{klz}
\end{eqnarray}
with $H_i \equiv ( \dot{a} /a, \dot{a} /a, \dot{a_z} /a_z)$
$\equiv (H_1,H_2=H_1, H_z)$ for $r, \theta$, and $z$ component
respectively.

 Given a Lagrangian
$L = \sqrt{g} {\cal L}=L(b(t), a((t), a_z(t))$, it can be shown
that
\begin{eqnarray}
L &=& { a^2 a_z \over \sqrt{B_1}} {\cal L} (R^{ti}_{\;\;\;tj},
R^{ij}_{\;\;\;kl}) = { a^2 a_z \over \sqrt{B_1}} {\cal L} (H_i,
\dot{H}_i, a^2)
\end{eqnarray}
The variational equations for this action can be shown to be:
\cite{kao06}
\begin{eqnarray} \label{key0}
{\cal L} +H_i ( {d \over dt} +3H )L^i &=& H_iL_i + \dot{H}_i L^i \\
{\cal L} + ( {d \over dt} +3H )^2 L^z &=& ( {d \over dt} +3H )L_z
\end{eqnarray}
Here $L_i \equiv \delta {\cal L} /\delta H_i$,  $L^i \equiv \delta
{\cal L} /\delta {\dot H}_i$, and $3H \equiv \sum_i H_i$. For
simplicity, we will write ${\cal L}$ as $L$ from now on in this
paper. As a result, the field equations can be written in a more
comprehensive form:
\begin{eqnarray} \label{key}
DL &\equiv&  L +H_i ( {d \over dt} +3H )L^i - H_iL_i - \dot{H}_i L^i =0\\
D_z L &\equiv& L + ( {d \over dt} +3H )^2 L^z - ( {d \over dt} +3H
)L_z =0  \label{zeq}
\end{eqnarray}

\section{higher derivative gravity model with a scalar field}

In this section, we will study the higher derivative induced
gravity model:
\begin{equation}
{ L} = - R - \alpha R^2 - \beta R^\mu_\nu R^\nu_\mu + \gamma
R^{\mu \nu}_{\;\;\; \beta \gamma} \, R^{\beta \gamma}_{\;\;\;
\sigma \rho} \, R^{\sigma \rho}_{\;\;\; \mu \nu} -{1 \over 2}
\partial_\mu \phi \partial^\mu \phi - V(\phi) \equiv   L_g +L_\phi
\end{equation}
with $L_g = -R-\alpha R^2 - \beta R^\mu_\nu R^\nu_\mu+ R^{\mu
\nu}_{\;\;\; \beta \gamma} \, R^{\beta \gamma}_{\;\;\; \sigma
\rho} \, R^{\sigma \rho}_{\;\;\; \mu \nu}$ and $ L_\phi = -{1
\over 2} \partial_\mu \phi \partial^\mu \phi - V(\phi)$ denoting
the pure gravity terms and the scalar field Lagrangian
respectively.

The corresponding Lagrangian can be shown to be:
\begin{eqnarray}&&
L= 2(2A+B+2C+D)- 4 \alpha \left[
4A^2+B^2+4C^2+D^2+4AB+8AC+4AD+4BC+2BD+4CD \right] \nonumber
\\ && - 2 \beta \left[3A^2+B^2+3C^2+D^2+2AB+2AC+2AD+2BC+2CD
\right]+ 8{\gamma } \left[2A^3+B^3+2C^3+D^3
\right]\nonumber \\
&&+{1 \over 2}\dot{\phi}^2-V(\phi)
\end{eqnarray}
with
\begin{eqnarray}
&& A=\dot{H}_1+H_1^2, \\
&& B= H_1^2 + {k \over a^2}, \\
&& C=H_1H_z, \\
&& D=\dot{H}_z+H_z^2 .
\end{eqnarray}
This Lagrangian can be shown to reproduce the de Sitter models
when we set $H_i \to H_0$ in the isotropic limit. The Friedmann
equation reads:
\begin{eqnarray} \label{keyphi}
 DL_g  = {1 \over 2}{\dot{\phi}}^2
+V(\phi)
\end{eqnarray}
for the induced gravity model. In addition, the scalar field
equation can be shown to be:
\begin{equation} \label{phieq}
\ddot{\phi} +3 H_0 \dot{\phi} +V'= 0.
\end{equation}
The leading order de Sitter solution with $\phi=\phi_0$ and
$H_i=H_0$ for all directions can be shown to be:
\begin{eqnarray} \label{Feq0}
V_0 &\equiv& V(\phi_0) = 6 [1-4\gamma H_0^4]H_0^2 .
\end{eqnarray}
Note that the coupling constants $\alpha$ and $\beta$ do not
affect the strength of inflation determined by the Hubble constant
$H_0$. This polynomial equation can be written as an cubic
equation in terms of the variable $x=H_0^2$:
\begin{equation}\label{fy}
f(x) \equiv x^3- { 1 \over 4\gamma} x+{V_0 \over 24 \gamma}=0 .
\end{equation}
It can be shown that the cubic polynomial $f(x)$ attains its
maximum and minimum at $x=x_M$ and $x=x_m$ respectively with $x_m=
-x_M= 1/[2 \sqrt{3 \gamma}]$. Plotting $C \equiv (\, x, \; y=f(x)
\;)$ as a curve $C$ on the $x$-$y$ plane, it is easy to show that
the curve $C$ intersects with the $y$-axis at the point $Y =( 0,
V_0/[24 \gamma])$.

If $f(y)$ attains its maximum and minimum at the points $M=(x_M,
f(x_M))$ and $N=(x_m, f(x_m))$ respectively, it can be shown that
$f(x=\pm x_m)= (V_0 \mp 2/\sqrt{3 \gamma})/[24 \gamma]$. Note that
the function $f(x) \to \pm \infty$ as $x \to \pm \infty$.
Therefore curve $C$ will have two intersection points $x_+=(x_+,
0)$ and $x_-=(x_-,0)$ with the positive $x$-axis if the the
minimum point $N$ locates at the quadrant IV, or equivalently,
$f(x_m)< 0$. This implies that the cubic equation $f(x)=0$ has two
positive roots if $3 \gamma V_0^2 < 4$. There is, however, only
one positive root when $x_+=x_-$, or equivalently, when $3 \gamma
V_0^2 = 4$.

Indeed, the cubic equation $f(x)=0$ can be solved by using the
triple angle formula of cosine function:
\begin{eqnarray}\label{yi}
\cos [3\theta] = 4 \cos^{3} \theta - 3 \cos \theta \; .
\end{eqnarray}
As a result, the solutions to the cubic function can be shown to
be
\begin{eqnarray}\label{xi}
x_1&=& -\left({1 \over 3 \gamma} \right)^{1/2} \cos {\theta_0
\over
3} \\
x_\pm &=& \left(  {1 \over 3 \gamma} \right)^{1/2} \cos \left ( {
\theta_0 \mp \pi \over 3}  \right ) . \label{xp}
\end{eqnarray}
with $\theta_0$ defined by the identity $\cos \theta_0 \equiv
\sqrt{ 3\gamma} \; V_0 /2 \le 1$. The notation $x_\pm$ is defined
according to its orientation with respect to the $x$-coordinate,
i.e. $x_- \le x_m \le x_+$. These solutions can be easily
converted to the well known solutions when $3 \gamma V_0^2 \ge 4$
remains valid.

 Note that $H_0=x_\pm^{-1/2}>0$ can have two
different choices as long as $3 \gamma V_0^2 < 4$ is satisfied.
These two solutions become identical to each other when $3 \gamma
V_0^2 = 4$, or equivalently, when $\cos \theta_0 =1$. It is also
straightforward to verify that $x_i$ given above does solve the
cubic polynomial equation $f(x)=0$ in consistent with the
constraint: $3 \gamma V_0^2 \le 4$.  In addition, $x_1$ can not be
a physical solution because it is negative.

\section{Stability of higher derivative inflationary solution}

Our universe could start out anisotropic and evolves to the
present highly isotropic state in the post inflationary era.
Therefore, a stable KS$/$FRW solution is necessary for any
physical model of our universe.

Given an effective action of the sort described by Eq. (\ref{V0}),
the scale factor $H_0$ is determined by the solution $H_0^2=x_\pm$
in the presence of the de Sitter solution $H_i=H_0$ and the static
condition $\phi=\phi_0$. In addition to these constraints, small
perturbations, $H_i=H_{i0}+ \delta H_i$ and $\phi=\phi_0+\delta
\phi$, against the background de Sitter solution $(H_{i0},
\phi_0)$ may also put a few more constraints to the system. This
perturbation will enable us to understand whether the background
solution is stable or not.

It can be shown that the perturbation equation of the Bianchi
models are identical to the perturbation equation of the FRW
models. Therefore, any inflationary solutions with a stable mode
and an unstable mode could provide us with a natural resolution to
problem of graceful exit. Such models will, however, also be
unstable against the anisotropic perturbations. Therefore, any
inflationary solutions with a stable mode and an unstable mode is
also negative to our search for a stable and isotropic
inflationary model. It will be shown shortly that the higher
derivative gravity theory with an inflaton scalar field could
hopefully resolve this problem all together.

In practice, perturbing the background de Sitter solution along
the $\delta H_i$ direction should be stable for at least a brief
moment such that around $60$ e-fold inflation can be induced
before the de Sitter phase collapses. And the resulting universe
should also be stable against isotropic and anisotropic
perturbations. In addition, the scalar field is expected to roll
slowly from the initial state $\phi=\phi_0$. Therefore, the
perturbation along the $\delta \phi$ direction is expected to be
unstable favoring the system for a natural mechanism of graceful
exit. Hence, we will try to study the stability equations of the
system for small perturbations against the de Sitter background
solutions in this section.

The first order perturbation equation for $DL$, with $H_i \to
H_0+\delta H_i$, can be shown to be \cite{kao06}:
\begin{eqnarray}\label{stable0}
\delta ( DL) &=& <H_i L^{ij} \delta  \ddot{H}_j> +3H <H_i L^{ij}
\delta \dot{H}_j> + \delta <H_i\dot{L}^i>+3H <(H_i L^i_j + L^j)
\delta
H_j> \nonumber \\
&& +<H_iL^i> \delta (3H) -<H_iL_{ij} \delta H_j>\label{stable1}
\end{eqnarray}
for any $DL$ defined by Eq. (\ref{key}) with all functions of
$H_i$ evaluated at some FRW background $H_i=H_0$. The notation $<
A_iB_i> \equiv \sum_{i=1,z} A_iB_i$ denotes the summation over
$i=1$ and $z$ for repeated indices. Note that we have absorbed the
information of $i=2$ into $i=1$ since they contributes equally to
the field equations in the KS type spaces. In addition, $L^{i}_{j}
\equiv \delta^2 L / \delta \dot{H}_{i} \delta H_j$ and similarly
for $L_{ij}$ and $L^{ij}$ with upper index $^i$ and lower index
$_j$ denoting variation with respect to $\dot{H}_i$ and $H_j$
respectively for convenience. In addition, perturbing Eq.
(\ref{zeq}) can also be shown to reproduce the Eq. (\ref{stable0})
in the de Sitter phase \cite{kao06}.

In addition, it can be shown that
\begin{eqnarray}
<H_i L^{i1}> &=& 2 <H_i L^{iz}  > ,\\
<H_i L^i_1> &=& 2 <H_i L^i_z>, \\
L^1 &=& 2 L^z ,\\
<H_iL_{i1}> &=& 2 <H_iL_{iz}>,
\end{eqnarray}
in the inflationary de Sitter background with $H_0=$ constant.
Therefore, the stability equations (\ref{stable1}) can be greatly
simplified. For convenience, we will define the operator ${\cal
D}_L$ as
\begin{eqnarray}
{\cal D}_L \delta H \equiv  <H_i L^{i1}> \delta  \ddot{H} +3H <H_i
L^{i1}> \delta \dot{H} +3H <H_i L^i_1 + L^1>  \delta H+2 <H_iL^i>
\delta H - <H_iL_{i1}> \delta H.
\end{eqnarray}
As a result, the stability equation (\ref{stable1}) becomes
\begin{equation}
\delta (DL)=  {\cal D}_L (\delta H_1+ \delta H_z/2)={3 \over
2}{\cal D}_L (\delta H)
\end{equation}
with $H=(2 H_1+H_z)/3$ as the average of $H_i$.

Hence the leading order perturbation equation in $\delta H$ and
$\delta \phi$ of the Friedmann equation can be shown to be:
\begin{equation}  \delta (DL_g) = {3 \over 2}{\cal D}_g \delta
H=V' \delta \phi
\end{equation}
with ${\cal D}_g \delta H \equiv {\cal D}_{L_g} \delta H $ as
short-handed notations. This equation can further be shown to be:
\begin{equation}
12 H_0 \left \{  2\left[ 6 \gamma H_0^2 -3\alpha -\beta \right]
(\delta \ddot{H} +3H_0 \delta \dot{H}) + \left [ 1-12 \gamma H_0^4
\right ]\delta H \right \} = V'_0 \delta \phi.
\end{equation}
Similarly, the leading perturbation of the scalar field equation
can be shown to be:
\begin{equation}
\delta \ddot{\phi} +3H_0 \delta \dot{\phi} +V''_0 \delta \phi =0
\end{equation}
The variational equation of $a_z$ can also be shown explicitly to
be redundant in the limit $H_i =H_0+\delta H_i$ and $\phi=\phi_0+
\delta \phi$ following the Bianchi identity.

Assuming that $\delta H=\exp[hH_0t] \delta H_0$ and $\delta \phi
=\exp[pH_0t] \delta \phi_0$ for some constants $h$ and $p$, one
can write above equations as:
\begin{eqnarray}
12 H_0 \left \{  2\left[ 6 \gamma H_0^2 -3\alpha -\beta
\right]H_0^2 ( h^2+3h) \delta H  + \left [ 1-12 \gamma H_0^4
\right ]\delta H \right \} &=& V'_0 \delta \phi. \\
H_0^2 ( p^2 + 3p + {V''_0 \over H_0^2} )\delta \phi &=&0 .
\end{eqnarray}
These equations are consistent when all coefficients vanish
simultaneously. This implies that $V'_0=0$ and
\begin{eqnarray}
h^2+3h + { 1-12\gamma H_0^4 \over 2H_0^2 \left [ 6 \gamma H_0^2
-3\alpha -\beta
\right ]} &=& 0, \\
( p^2 + 3p + {V''_0 \over H_0^2} )&=&0 .
\end{eqnarray}
As a result, the solutions to $p$ are $p=p_\pm= -3/2 \pm
\sqrt{9-4V''_0/H_0^2 \;} /2$. Since $V'_0=0$, the scalar field
equation can be solved in the de Sitter background $H_i=H_0$.
Indeed, the solution to the equation $\ddot{\phi} +3 H_0
\dot{\phi} \sim 0$ is:
\begin{equation}
\phi \sim \phi_0 + {\dot{\phi}_0 \over 3 H_0}[1 - \exp (- 3 H_0 t)
] \label{phit} .
\end{equation}
This result indicates that the scalar field does change very
slowly similar to the slow roll-over assumption in various scalar
field models.

An appropriate effective spontaneously symmetry breaking (SSB)
potential $V$ of the following form
\begin{equation} \label{V0}
V(\phi)= { \lambda \over 4} (\phi^2-v^2)^2 +V_m
\end{equation}
with arbitrary coupling constant $\lambda$  can be shown to be a
good candidate for our physical universe. Here $V_m$ is a small
cosmological constant dressing the SSB potential. The local
extremum of this effective potential can be shown to be $\phi=0$
(local maximum) and $\phi=v$ (local minimum). We expect that the
scalar field starts off from the initial state $\phi =0$ and rolls
slowly down to its local minimum locates at $\phi=v$.

When the scalar field eventually rolls down to the minimum of $V$
at $\phi=v$, the system will oscillate around this local minimum
with a friction term related to the effective Hubble constant
$H_m$ at this stage. Reheating process is expected to take away
the kinetic energy of the scalar field. The scalar field will
eventually becomes a constant background field losing all its
dynamics.

The value of $H_0$ can be chosen to induce enough inflation for a
brief moment as long as the slowly changing scalar field remains
close to the initial state $\phi=\phi_0=0$. The de Sitter phase
will hence remain valid and drive the inflationary process for a
brief moment determined by the decaying speed of the scalar field.

In addition, at the initial stage,
\begin{equation}
p_\pm=-3/2 \pm \sqrt{9+ 4\lambda v^2/H_0^2 \;} /2
\end{equation}
for the SSB $\phi^4$ potential model. Therefore, it is easy to
prove that $p_+>0$ and $p_-<0$ indicating an unstable mode and a
stable mode do exist when we perturb the scalar field in this
model. Hence the system will be unstable against the perturbation
of the scalar field consistent with the slowly changing scalar
field solution (\ref{phit}).

On the other hand, the solutions to the $h$-equation are $h=h_\pm=
-3(1 \pm \delta_2)/2$ with
\begin{equation}
\delta_2^2=1+ 2 ( 12\gamma H_0^4-1)/\{9 H_0^2[ 6 \gamma H_0^2
-3\alpha -\beta ]\}
\end{equation}
indicating that $h_+<0$. Therefore, the $h$-perturbation can have
two stable modes only when $h_-<0$ or $\delta_2^2 <0 $. This
implies that either
\begin{eqnarray} \label{gammaa}
&(a)&  1< 12 \gamma H_0^4 < 2(3 \alpha+\beta)H_0^2
\\ {\rm or} \;\;\; &(b)& 2 (3 \alpha+\beta)H_0^2 < 12 \gamma H_0^4
<1 \label{gammab}
\end{eqnarray}
has to hold. Let us define the function $\Gamma_\pm \equiv 12
\gamma H_\pm^4-1$ with $H^2_\pm \equiv x_\pm= [\cos (\theta \mp
\pi)/3] /\sqrt{3 \gamma}$ for arbitrary $\theta$. $\Gamma_\pm$
function can be written as
\begin{equation} \label{Gamma}
\Gamma_\pm = \pm 4 \sin { \theta \over 3} \cos [{\theta \over
3}\mp {\pi \over 6}].
\end{equation}
Therefore, it is easy to show that $\Gamma_+ \ge 0$ and $\Gamma_-
\le 0$ for all $0\le \theta \le \pi/2$. As a result, solution
$x_+$ is consistent only with the constraint (a) in Eq.
(\ref{gammaa}). Similarly, the solution $x_-$ is consistent only
with the constraint (b) in Eq. (\ref{gammab}).

In addition, the equation for the perturbation on anisotropic KS
type space and isotropic FRW space are the same. Therefore, the
stable $h$-perturbation also ensures that isotropy of the de
Sitter space can be made stable against any small anisotropic
perturbation $H_i=H_0 +\delta H_i$.

Once the scalar field falls close to its local minimum at
$\phi=v$, the scalar field will start to oscillate around this
local minimum. The Hubble parameter at this stage will read
$H_i=H_m$ with $H_m$ given by the solution to the leading order
Friedmann equation at this stage with $\phi=v$ and $H_i=H_m$:
\begin{eqnarray} \label{Feqm}
V_m &\equiv& V(v) = 6 [1-4\gamma H_m^4]H_m^2.
\end{eqnarray}
The cubic equation in $x=H_m^2$ also reads:
\begin{equation}\label{Hm}
g(x)=x^3- { 1 \over 4\gamma} x+{V_m \over 24 \gamma}=
(x-x_2)(x-x_{2+})(y-x_{2-})=0,
\end{equation}
with
\begin{eqnarray}\label{xm}
x_2&=& -\left({1 \over 3 \gamma} \right)^{1/2} \cos {\theta_m
\over
3}, \\
x_{2\pm} &=& \left(  {1 \over 3 \gamma} \right)^{1/2} \cos \left (
{ \theta_m \mp \pi \over 3}  \right ) . \label{xpm}
\end{eqnarray}
and $\theta_m$ defined by the identity $\cos \theta_m \equiv
\sqrt{ 3\gamma} \; V_m /2 \le 1$. Note again that only positive
roots $x_{2\pm}$ can be physical solutions to $H_m^2$. In
addition, $H_0$ should be much larger than $H_m$ in order to
generate inflation. Moreover, any physical solution should also
respect the fact that $V_0 > V_m$ such that these solutions can be
made consistent with the SSB potential discussed here.

Indeed, the function $y_\pm \equiv (3 \gamma)^{1/2} x_\pm \equiv
\cos { (\theta \mp \pi)/3 } $ (with $\cos \theta \equiv \sqrt{
3\gamma} \; V/2$) can be shown to be a monotonically
increasing/decreasing function in $\theta$. Therefore, when
$\theta$ decreases, corresponding to the increasing of $V$,
$y_\pm$ will decrease/increase accordingly. Hence the relations
$H_0^2 > H_m^2$ and $V_0 > V_m$ are consistent if we choose the
set of solutions as ($x_-$,  $x_{2-}$). Therefore, the Hubble
constant $H_m$ at the final stage can be chosen to be much smaller
than the Hubble constant $H_0$ at the inflationary era with proper
choice of coupling constants.

We can also choose the set of solutions as $(x_+, x_{2-})$. In any
case, the initial condition $\theta_0 \sim 0$ and the final
condition $\theta_m \sim \pi/2$ can be adjusted with the proper
choice of $V_0$ and $V_m$. Note that $\theta_0$ and $\theta_m$ can
be chosen to be close to $0$ and $\pi/2$ respectively.

In addition, the first order perturbation equations also have two
modes: $p=p_\pm =-3/2 \pm [9-4V''_m/H_m^2 \;]^{1/2} /2=-3/2 \pm
[9- 8\lambda v^2/H_m^2 \;]^{1/2} /2$ for the SSB $\phi^4$
potential model (\ref{V0}). Stability is also ensured even if the
discriminator becomes imaginary. Therefore, there is no further
constraints to be imposed on the parameters from the $p$-mode
perturbation.

There are also two modes for the $h$-equation: $h=h_\pm= -3(1 \pm
\delta_3)/2$ with
\begin{equation}
\delta_3^2=1+ 2 ( 12\gamma H_m^4-1)/\{9 H_m^2[ 6 \gamma H_m^2
-3\alpha -\beta ]\}.
\end{equation}
Both $h$-modes must be stable for the existence of a stable
isotropic de Sitter space in the post-inflationary era. This
implies that either
\begin{eqnarray}
&(a)&1< 12 \gamma H_m^4 < 2(3 \alpha+\beta)H_m^2
\\ {\rm or} \;\;\; &(b)&  2 (3 \alpha+\beta)H_m^2 < 12 \gamma H_m^4 <1
\end{eqnarray}
has to hold. Note that constraint (a) and constraint (b) are
organized in a similar way for both $H_0$ and $H_m$. This will
make our analysis and comparison more easily.

If we take the set of solutions as $(x_-, x_{2-})$ consistent with
$H_0 \gg H_m$, both $H_0$ and $H_m$ should obey the constraint
(b). Indeed, Eq.s (\ref{gammaa}-\ref{gammab}) imply that $12
\gamma H_0^4 \le 1$ and $12 \gamma H_m^4 \le 1$. Similarly, if we
take the set of solutions as $(x_+, x_{2-})$, it can be shown that
$12 \gamma H_0^4 \ge 1$ and $12 \gamma H_m^4 \le 1$. Therefore, we
should choose constraint (a) for $H_0$ and constraint (b) for
$H_m$. Explicit calculation shows, however, that this will lead to
inconsistent constraint on the factor $ k_1 \equiv (3 \alpha
+\beta)/(6 \gamma)$. Indeed, the inequality $12 \gamma H_0^4 < 2(3
\alpha+\beta)H_0^2$ implies that $H_0^2 < k_1$, but the inequality
$12 \gamma H_m^4 > 2(3 \alpha+\beta)H_m^2$ implies instead that
$H_m^2 > k_1$. This is inconsistent with the fact that $H_0> H_m$.
Therefore this case can be excluded. As a result, the only
consistent set of solutions for $H_0$ and $H_m$ are derived from
the set of solutions $(x_-, x_{2-})$.

Consider the special case where $V_m=0$. The cubic equation
becomes $x(4 \gamma x^2-1)=0$. This implies that $4\gamma
H_m^4=1$. Or equivalently, this solution can also be obtained by
taking the limit $\theta_m = \pi/2$ in Eq.s (\ref{xm}, \ref{xpm}).
Therefore, the solution to $h$ equation is $h=h_\pm= -3(1 \pm
\delta_m)/2$ with
\begin{equation}
\delta_m = \sqrt{1+ 8/[27-9 (6 \alpha+2 \beta)H_m^2 ]}.
\end{equation}
The situation at this stage is similar to the case of pure gravity
model with a vanishing cosmological constant \cite{kao06}.

For the special case $V_m=0$, we also expect that these two
anisotropic perturbation modes against the isotropic background
solution $H_i=H_m +\delta H_i$ are both stable modes. As a result,
$\Delta_{\pm} \equiv - [3H_0t / 2] [1 \pm \delta_m]<0$ can be
shown to be the necessary condition, implying that $\delta_m < 1$
or, equivalently, $ (3 \alpha +\beta) H_m^2
> 35/18$. In particular, both modes becomes oscillatory triangular
functions when $35/18 > (3 \alpha +\beta) H_m^2 > 3/2$. In both
cases, the de Sitter space will remain a stable background as the
universe evolves.

\section{conclusion}

The existence of a stable de Sitter background is closely related
to the choices of the coupling constants. The pure higher
derivative gravity model with quadratic and cubic interactions
\cite{kao06} admits an inflationary solution with a constant
Hubble parameter. Proper choices of the coupling constants allow
the de Sitter phase to admit one stable mode and one unstable mode
for the anisotropic perturbation.

The stable mode favors a strong inflationary period and the
unstable mode provides a natural mechanism for the graceful exit
process. It is also found that the perturbation against the
isotropic FRW background space and the perturbation against the
anisotropic KS type background space obey the same perturbation
equations. This is true for both pure and induced gravity models.
As a result, the unstable mode in pure gravity model also means
that the isotropic de Sitter background is unstable against
anisotropic perturbations. Therefore, small anisotropic could be
generated during the de Sitter phase for pure gravity model.

We have shown that, for gravity models with an additional inflaton
scalar field, stable mode for perturbations along the anisotropic
$\delta H_i$ directions do exist with proper constraints to be
imposed on the coupling constants. In addition, another unstable
mode also exist for the perturbation against the scalar field
background $\phi_0$ with proper constraints. Therefore, the de
Sitter background can remain stable and isotropic during the
inflationary process for this inflaton gravity models.

Explicit model with a spontaneously symmetry breaking $\phi^4$
potential is presented as an example. Proper constraints are
derived for reference. It is shown that the these conditions must
hold for the initial de Sitter phase solutions in order to admit a
constant background solution during the inflationary phase for the
system.

Once the scalar field rolls down to the minimum at $\phi=v$, the
system picks up a small Hubble constant $H_m$ characterized by the
small cosmological constant $V_m$. We also show that the resulting
de Sitter space can be a stable final state with proper choices of
coupling constants. Therefore, the scalar-tensor theory proposed
in this paper is capable of inducing inflation and maintaining a
stable FRW de Sitter background space all together.

In summary, we have shown that a stable mode for (an)isotropic
perturbation against the de Sitter background does exist for the
gravity model with a scalar field. The problem of graceful exit
can rely on the unstable mode of the scalar field perturbation
against the constant phase $\phi_0$.

It is also found that the quadratic terms will not affect the the
inflationary solution characterized by the Hubble parameter $H_0$
and $H_m$. These quadratic terms play, however, critical role in
the stability of the de Sitter background. In addition, it is also
interesting to find that their coupling constants $\alpha$ and
$\beta$ always show up as a linear combination of $3\alpha +
\beta$ in these stability equations. Implications of these
constraints deserve more attention for the applications to the
inflationary models.

\section*{Acknowledgments}

This work is supported in part by the National Science Council of
Taiwan.

\end{document}